%% file: paper.tex
\documentclass[runningheads]{llncs}
\usepackage{graphicx}
\usepackage{url,xcolor}
\usepackage{booktabs}
\usepackage{amsmath,amssymb}
\usepackage{amsfonts,mathrsfs}
\usepackage{bm} 
\usepackage{arydshln} 
\usepackage{cancel}
\usepackage[scale=0.8]{geometry}

\newcommand{\slfrac}[2]{\left.#1\middle/#2\right.}

\newcommand{\bmat}[1]{\begingroup%
	\renewcommand*{\arraystretch}{1.1}
	\begin{bmatrix} #1 \end{bmatrix}
	\endgroup}
\begin{document}
\title{Perfusion Quantification from Endoscopic Videos: Learning to Read Tumour Signatures\thanks{%
Submitted to 23rd International Conference on Medical Image Computing \& Computer Assisted Intervention (MICCAI 2020).\protect\newline
This work was partially supported by the Disruptive Technology Innovation Fund, Ireland, project code DTIF2018\_240 CA.
}}
%
\titlerunning{Learning to Read Tumour Signatures}
%
%
\author{Sergiy Zhuk\inst{1} \and Jonathan Epperlein\inst{1} \and Rahul Nair\inst{1} \and Seshu Thirupati\inst{1} \and Pol Mac Aonghusa\inst{1} \and Ronan Cahill\inst{2} \and Donal O'Shea\inst{3}}
\authorrunning{Zhuk, Epperlein et al.}
%
\institute{IBM Research - Europe, Dublin, Ireland \email{\{sergiy.zhuk,jpepperlein,rahul.nair,seshutir,aonghusa\}@ie.ibm.com}\and University College Dublin, Dublin, Ireland \email{ronan.cahill@ucd.ie} \and Royal College of Surgeons, Dublin, Ireland \email{donalfoshea@rcsi.ie}}

\maketitle              
\begin{abstract}
  Intra-operative identification of malignant versus benign or healthy tissue is a major challenge in fluorescence guided cancer surgery. We propose a perfusion quantification method for computer-aided interpretation of subtle differences in dynamic perfusion patterns which can be used to distinguish between normal tissue and benign or malignant tumors intra-operatively in real-time by using multispectral endoscopic videos. The method exploits the fact that vasculature arising from cancer angiogenesis gives tumors differing perfusion patterns from the surrounding tissue, and defines a signature of tumor which could be used to differentiate tumors from normal tissues. Experimental evaluation of our method on a cohort of colorectal cancer surgery endoscopic videos suggests that the proposed tumor signature is able to successfully discriminate between healthy, cancerous and benign tissue with 95\% accuracy.
\keywords{Perfusion quantification\and bio-physical modeling \and Explainable features design \and Tissue classification \and Cancer.}
\end{abstract}
%
%

\section{Introduction}
\label{sec:introduction}
\input{intro}

\section{Methods}
\label{sec:decis-supp-syst}

\subsubsection{Data preprocessing: Fluorescence intensity estimation.}
\label{sec:endosc-moti-comp}
\input{timeseries}

\subsection{Parametric bio-physical models}
\label{sec:biophys-models-fluor}
\input{biomodels}

\section{Experimental validation: tissue classification}
\label{sec:exper-valid-tiss}

\subsection{ROI classification}%
\label{sec:feat-distr-healthy}
\input{classifier}

\subsection{Dataset and evaluation results }
\label{sec:datas-eval-results}

\label{sec:exper-valid}
\input{experiment}

\section{Conclusions}
\label{sec:conclusions}
We propose a method, based on \emph{bio-physical modeling} of in-vivo perfusion, that characterizes dynamic perfusion patterns by a compact signature, a vector of twelve biophysical features which could be used to differentiate tumours from normal tissues.  For validation of the proposed signature we implemented an experimental framework combining computer vision with our bio-physical model and machine learning (ML) using readily available open source tools. Experiments on a corpus of 20 colorectal cancer multispectral endoscopic videos captured during surgical procedures, showed that the generated signature of perfusion parameters is discriminant for benign, malignant and normal tissues.

Overall, experimental results suggest that our approach can reproduce the expert judgement of both a surgeon intra-operatively and of post-operative pathology analysis.
Our results suggest that our approach is a promising step towards fully automated intra-operative interpretation of tissue enabling future research in areas such as tumor delineation and tissue structure modeling. An immediate research priority is to further test and refine our models by scaling up the collection, processing and classification of videos to include more collaborators, procedures and applications. In the longer term integration with hardware platforms for robotic surgery is also a promising avenue for future research.

\bibliographystyle{splncs04}
\bibliography{../DTIF}
\end{document}

%% file: intro.tex
\begin{figure}[tb]
   \centering
   \includegraphics[width=.8\columnwidth]{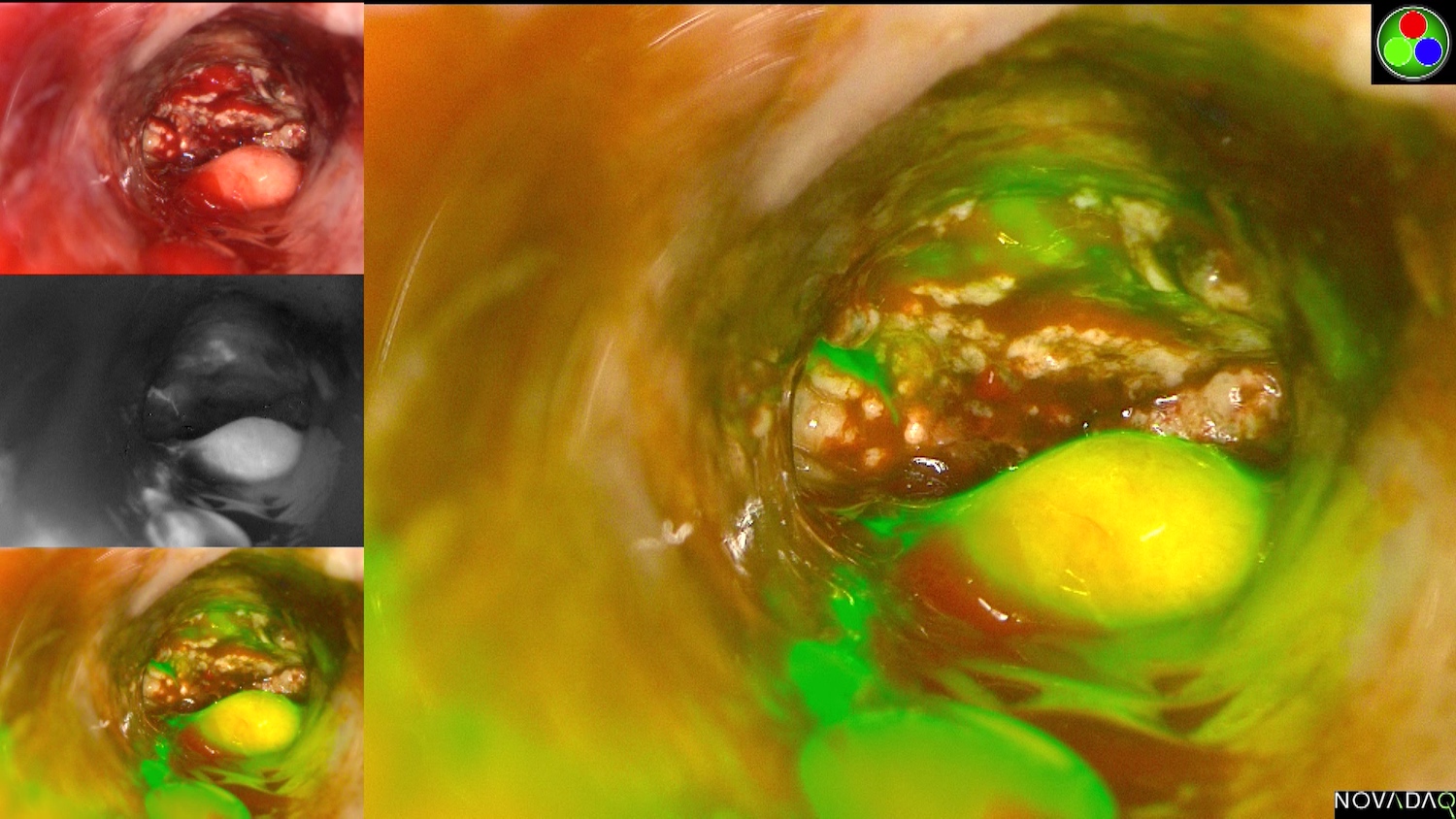}
   \caption{Left:~visible light frame (top); NIR frame (center); brightness from NIR frame is used as green channel in vis.light frame (bottom); large panel cycling between 3 panels to the left.}
   \label{fig:frame1}
 \end{figure}

Quantification of perfusion\footnote{Perfusion is the passage of fluid through the circulatory or lymphatic system to a capillary bed in tissue.} using a fluorescent dye, such as Indocyanine Green (ICG), has become an important aid to decision making during surgical procedures~\cite{Boni2015,Huh2019}. ICG is currently utilized in fluorescence-guided surgery for identification of solid tumors~\cite{schaafsma2011clinical}, verification of adequate perfusion prior to anastomosis during colorectal resection~\cite{ALES4789}, lymph node mapping during lymphadenectomy~\cite{Vesale443}, and identification of biliary tract anatomy during colectomy~\cite{doi:10.1177/1553350617690309}. Visible and near infra-red (NIR) light sources are widely available in laparoscopic/endoscopic cameras, providing high-resolution, multispectral video of blood flow in tissue and organs. An example frame captured by a \emph{Stryker PINPOINT\textsuperscript{TM}} clinical camera is shown in Figure~\ref{fig:frame1}.

Intra-operative identification of malignant versus benign or healthy tissue is a major challenge of cancer surgery. A surgeon typically uses a combination of visual and tactile evidence to identify tissue during a surgical procedure. ICG has been observed to accumulate in cancers, \cite{veys2018icg}. In practice, however, consistent intra-operative detection of sometimes subtle and complex differences in ICG perfusion has proven challenging. Intra-operative interpretation requires a surgeon to track spatial and temporal fluorescence intensity patterns simultaneously over several minutes, and it can be challenging to distinguish variations in relative fluorescence intensity due to confusing factors such as inflammation~\cite{HOLT2014}.
We hypothesize that observation of differences in structure of vasculature and perfusion patterns using ICG-enhanced fluorescence could be used to differentiate between benign,  malignant, and healthy tissue, and that \emph{perfusion patterns characterized by ICG inflow and outflow can serve as a marker to identify most of the benign and malignant tumors intra-operatively}~\cite{HOLT2014}.

It seems natural to ask, therefore, whether \emph{computer assisted interpretation} of perfusion could assist with interpretation of perfusion by approximating the judgement of a surgeon with high probability?
In addressing this question we propose a method, as our key contribution, based on \emph{bio-physical modeling} of in-vivo perfusion.  Our model characterizes dynamic perfusion patterns by (i) estimating time-series of ICG fluorescence intensities, representing ICG inflow and outflow within a region of the tissue, from multispectral videos, and (ii) fitting parameters of a bio-physical perfusion model to the estimated time-series to create a compact signature of perfusion consisting of a vector of bio-physical features of ICG inflow and outflow, (iii) showing experimentally that the generated signature of perfusion parameters is discriminant for benign, malignant and normal tissue using traditional machine learning (ML) techniques.
Our method is agnostic to camera technologies employed, relying only on the availability of multispectral video signals from a laparoscopic or endoscopic fluorescence imaging camera. The parameters of our bio-physical model are readily interpretable, and the derived signature characterizes pharmacokinetics of blood-ICG admixtures in tissue by a vector of features described below (see~Table~\ref{tab-features}).

We perform experimental validation on a corpus of 20 colorectal cancer multispectral endoscopic videos captured during surgical procedures.
An experimental framework is implemented combining computer vision with a bio-physical model and machine learning. By estimating time-series of ICG intensities for a number of randomly selected Regions of Interest (ROIs) we fit parameters from our bio-physical model to time-series of ICG intensities and derive a signature for each ROI. The derived bio-physical signatures are subsequently used as input features for standard supervised classification methods to attempt to differentiate normal ROIs from suspicious (benign or malignant tumors).
Experiments show that our approach can match the intra-operative interpretation of an expert surgeon with 86\% accuracy for ROI-based correctness, 
and 95\% accuracy for patient-based correctness (i.e., compared with  subsequent post-operative pathology findings on excised tissue) with 100\% sensitivity (percentage of correctly detected cancer) and 92\% specificity (percentage of correctly detected normal tissue).

Our choice of traditional ML technologies is deliberate, and intended to emphasize that high quality computer assisted interpretation in real-time can be provided even for limited-size datasets by combining basic ML tools with bio-physical models of perfusion. We suggest our approach represents a valuable step towards computer assisted automation of intra-operative interpretation of tissue.

\begin{figure}[tb]
  \centering
  	\includegraphics[width=\textwidth]{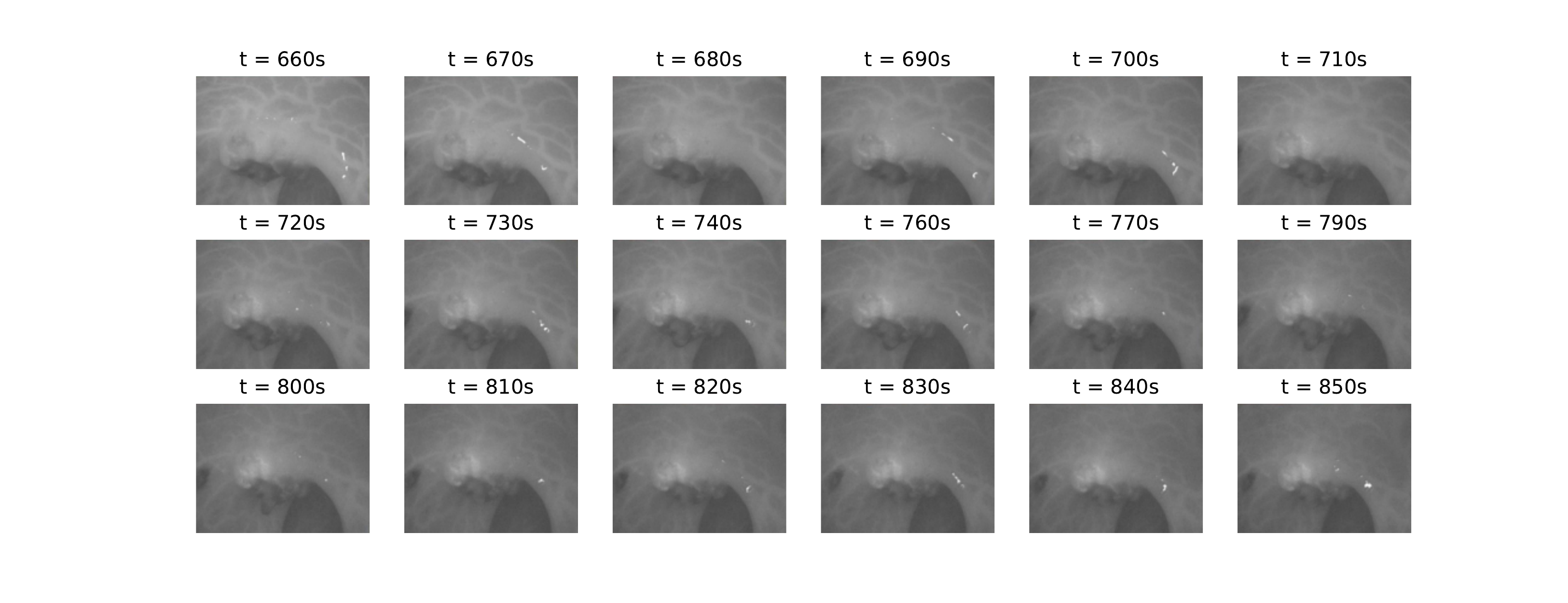}
	\caption{Perceptual hashing algorithm~\cite{idealods2019imagededup,cmc.2020.07421} applied on one of the perfusion videos used in this study. Selected images till the end of the perfusion $t=860$s are near duplicates of the first frame.}
\label{fig:intro:image_duplication}
    \end{figure}%

\paragraph{\bf Related work}
Image-based cancer detection methods have received significant attention with the advent of advanced fluorescence imaging systems and progress in image processing. We consider methods under three categories:
(1) image classification using deep learning,
(2) observation by the surgeon of perfusion in the colon
and, (3) quantitative analysis of bio-physical models based on perfusion patterns.

Deep learning 
has been applied for cancer screening using images from mammography (\cite{mckinney2020international}) and histology (\cite{shapcott2019deep}). Deep learning methods typically require significant numbers of labelled training examples for each class to estimate model parameters. The cancer screening study (\cite{mckinney2020international}) for example, required several tens of thousands of mammograms with known pathology. For image-based classification, several deep learning architectures have been proposed. Examples include robust recurrent convolutional neural networks and ensembles of deep learners that include pre-trained networks.


In contrast, the approach taken here involves feature engineering based on bio-physical models combined with transitional ML classifiers. This approach is better suited when training data is limited. Since we are primarily interested in perfusion dynamics, i.e.\ the evolution of fluorescence in specific tissue over time, the task is one of video classification. This task cannot be readily translated to an image classification problem for several reasons, precluding the use of existing methods.
\emph{Although containing many individual images, each surgical video represents only a few sufficiently distinct images and results in significant duplication when used as a training set}. As shown in Fig.~\ref{fig:intro:image_duplication}, for one of the videos used in this study, most of the frames of the NIR video are near duplicates when compared using perceptual hashing~\cite{idealods2019imagededup,cmc.2020.07421}.
Attempting to augment the training set by, for example, (i) tiling video's initial frame with many ROIs (see Fig.~\ref{fig:p18}
for an example of selecting ROIs) and creating a new video for each ROI by extracting the corresponding ROIs from every subsequent frame,
or (ii) extracting short time sequences of entire frames, or (iii) combining (i) and (ii) to increase the amount of data, is still observed to result in significant duplication for the training set in our experiments.

Observation of perfusion patterns by surgeons is common practice for anastomoses~\cite{Boni2015,foppa2014indocyanine}. However, the changes in the perfusion needed to discriminate cancer are not easily detected by visual observation (Fig.~\ref{fig:intro:image_duplication})~\cite{daskalaki2015fluorescence,james2015fluorescence,son2019quantitative}.

Differences in vasculature are used in~\cite{jayender2013statistical} to build a statistical hidden Markov model of perfusion from DCE-MRI images for breast cancer. A key difference between the approach in~\cite{jayender2013statistical} and the multispectral NIR video used here is that the NIR video has a much higher sampling rate (10 frames per second for both NIR and visible light) in comparison with DCE-MRI images in~\cite{jayender2013statistical} (taken every 90 seconds).

Mathematical models describing the bio-physics of perfusion provide an alternative method for detecting cancerous ROIs. The premise of this method is that differences in ICG inflow during a relatively short timeframe of minutes after injection known as the \emph{wash-in} phase, and ICG outflow during the beginning of the venous outflow, termed the \emph{wash-out} phase, can serve as a marker for most benign and malignant tumors. Indeed, cancer angiogenesis has been recognized to lead to abnormal vasculature , characterized by higher internal pressure 
and a discontinuous basement membrane, in comparison with normal, healthy tissue~\cite{nishida2006angiogenesis,de2017microenvironmental}. Hence, ICG inflow over a \emph{wash-in} phase could already disclose valuable discriminative signatures. In addition to this, it was noted that most of the malignant tumors are characterized by ``increased interstitial pressure from leaky vessels'' and ``relative absence of intra-tumoral lymphatic vessels''~\cite{folkman1995clinical} which suggests that fluid ``could pool'' in the malignant tumor over time causing the retention of the ``tracer'', thus demonstrating differences of ICG outflow during \emph{wash-out} phase. Finally, benign tumors are known to exhibit slower blood vessel growth, while malignant tumors are characterized by more rapid blood vessel growth~\cite{brustmann1997relevance} implying that ICG inflow/outflow patterns for benign and malignant tumors are different too.

Perfusion quantification based on estimating time-series of ICG fluorescence intensities, and then extracting a number of so called \emph{time-to-peak} (TTP) features directly from the time-series is well represented in the surgical literature. TTP-features have direct physical meaning (for example,  Fig.~\ref{fig:moson} in Section~\ref{sec:biophys-models-fluor}), and have been successfully applied to perfusion quantification on animals~\cite{diana2014enhanced}, and for predicting anastomotic complications after laparoscopic colorectal surgery~\cite{son2019quantitative}.

\emph{Biological compartment models},~\cite{vilanova2017mathematical,choi2011dynamic}, model the dynamics of ICG inflow to a small compartment of the tissue during the initial \emph{wash-in} phase, and ICG outflow during the subsequent \emph{wash-out} phase. As a typical example, 
 \cite{gurfinkel2000pharmacokinetics} model the ICG time-series as a sum of two exponentials. 
In~\cite{choi2011dynamic}, a slightly more general model 
using arterial, capillaries and tissue/extravascular compartments is used. These models were successfully tested on animals.

The bio-physical model proposed here generalizes models proposed in~\cite{choi2011dynamic,gurfinkel2000pharmacokinetics} by modeling ICG intensity time-series as a response of a generic second-order linear system with exponential input, which is a sum of one real and two complex exponentials, to allow for oscillating behaviors observed in ICG time-series estimated from videos of human tissue perfusion (e.g., Fig.~\ref{fig:frame1} and Fig.~\ref{fig:fits}).
The coefficients and exponents of these exponential terms form a set of features which we will call \emph{3EXP}.

To exploit the natural synergy between surgical approaches to perfusion quantification and bio-physics we define a \emph{tumor signature} as a combination of \emph{3EXP} and \emph{TTP} features. Combining features in this way results in significant improvements in accuracy over predictions obtained by using \emph{3EXP} or \emph{TTP} features separately taken in our experiments (see Section~\ref{sec:datas-eval-results}). The latter show that \emph{TTP} features slightly outperform \emph{3EXP} when comparing individual ROIs, while accuracy of patient-based comparison is significantly higher for \emph{3EXP} features. Most importantly, the combination of \emph{3EXP and TTP} features taken together are synergistic: 95\% accuracy for patient-based correctness (up by 10 percentage points) with 100\% sensitivity (up by 23 percentage points) and 92\% specificity. To the best of our knowledge this is the first time a combination of a generalized compartment model with TTP-features has been successfully applied to provide a discriminative signature of tumors in human tissue.


%% file: timeseries.tex
\begin{figure}[htb]
\parbox[t][][t]{0.99\columnwidth}{
  \centering
  \includegraphics[width=.85\columnwidth]{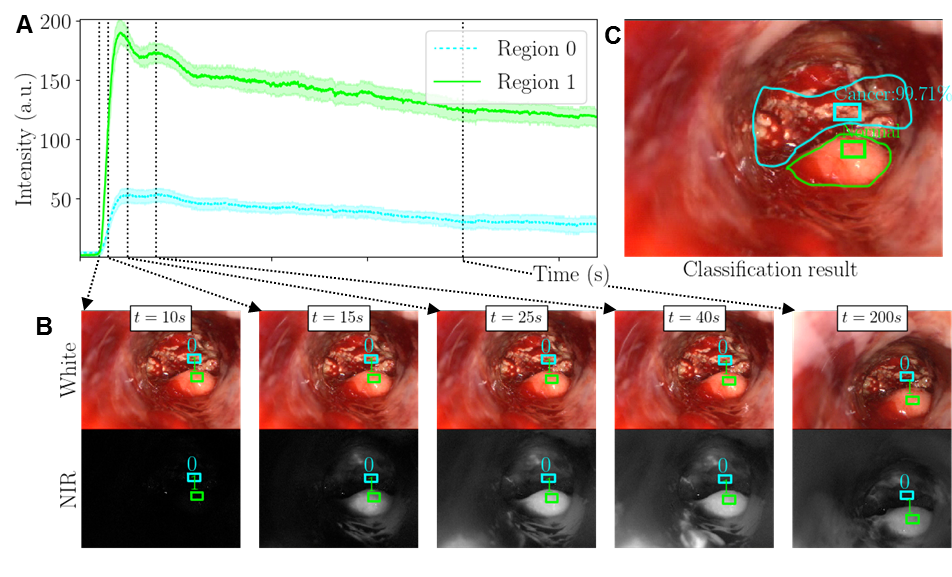}
	\caption{
	Panel A: NIR-intensity time-series for 300 seconds post ICG administration for two ROIs; blue trace $I_{0}(t)$ (ROI 0) and green trace $I_{1}(t)$ (ROI 1). At time instant $t$ the value $I_{i}(t)$ equals the mean of the intensity taken over ROI $i$, $i=0,1$, and the bands denote $\pm$ one standard deviation around that mean. Panel B: White -- visible light video sequence, NIR -- NIR light sequence used for NIR-intensity estimation. Panel C: ROI with surgical team annotation, and a classification results showing ROIs correctly classified as normal (green) and cancer (light blue).}
	\label{fig:p18}
      } \vspace{-.1cm}
    \end{figure}

The data sources here are composed of multispectral endoscopic videos (e.g.~Fig.~\ref{fig:frame1}). It is noted in~\cite{Benson_1978} that for ICG concentrations used in the surgical procedures considered here, \emph{fluorescence intensity at the peak wavelengths is proportional to the ICG concentration}, consequently we use the \emph{NIR intensity} extracted from the NIR-channel video frames (see~Fig.~\ref{fig:p18}) as a proxy for the ICG concentration in the tissue.

The initial frame of each video has areas of suspicious and normal areas identified by a surgical team (see Fig.~\ref{fig:p18}, panel C). We randomly select ROIs within those areas, e.g.\ ROI 0 and 1 in Fig.~\ref{fig:p18}. After injection of the NIR dye, the \emph{NIR intensity} within each of ROIs is extracted from the NIR video stream for as long as the ROI stays visible in the field of view. Data collection is straightforward only if the camera and the tissue within field of view do not move. For inter-operative surgical use, the camera is handheld and tissue contracts and expands, making acquisition very challenging. For example, ROI 0 in Fig.~\ref{fig:p18} between time $t=1s$ and $t=300s$ shows considerable drift.%
As this is typically the case in colorectal surgical procedures,  motion stabilization techniques~\cite{selka2014fluorescence} are required to compensate for motion during collection of time-series. In what follows we assume that the data collection has already taken place. 

The outcome of data preprocessing is a set of time-series estimating temporal evolution of the NIR intensity for each pixel contained in each ROI. The NIR intensity in each ROI is further aggregated by taking the mean brightness across the pixels in every ROI, and measure the variation by standard deviation as depicted in Fig.~\ref{fig:p18}, panel A. After aggregating the brightness across all the pixels within a ROI for each time-step, we get a time-series of intensities for each ROI; let $I_{p,r}(t)$ denote the aggregated intensity of ROI $r$ in patient $p$ at time $t$. The time axis is $t\in\{0,\delta_t, 2\delta_t, \dots, T_{p,r}\}$, where  $\delta_t$ is dictated by the frame rate of the imaging equipment (here, $\delta_t=0.1s$) and $T_{p,r}$ denotes the time at which either ROI $r$ was lost or tracking was eventually terminated.

%% file: biomodels.tex
To create bio-physical signatures we parametrize time-series of NIR intensities $I_{p,r}(t)$ as follows:
\begin{equation}
  \label{eq:ICGmodel}
y(t;\tau,D,K,\tau_i,\theta,y_{dc}) = y_{\exp}(t-\theta;\tau,D,K,\tau_i) \cdot H(t-\theta) + y_{dc}
\end{equation}%
\noindent where $y_{\exp}(t;\tau,D,K,\tau_i)$ is the response of a linear time-invariant second-order system to an exponential input~(see e.g.~\cite{phillipsharbor2000}), i.e.\ $y_{\exp}$ solves the differential equation
\begin{equation}
	\tau^2 \ddot{y}(t) + 2D\tau \dot{y}(t) + y(t) = K e^{-t/\tau_i} \label{eq:2ndorder},
\end{equation}%
\noindent with  zero initial conditions, $y(0)=\dot{y}(0)=0$; $\tau$ is the \textbf{time constant}, and $D$ is the \textbf{damping}, which together govern the speed of the response and whether there are oscillations; $K$ is known as the \textbf{gain} and responsible for the amplitude of the response, and $\tau^{-1}_i$ is the \textbf{input decay rate} which determines the rate of decay for large  $t$.~
Finally, $\theta$ in~\eqref{eq:ICGmodel} is a time delay, which accounts for the time it takes from the start of the video until the ICG reaches the imaged tissue, whereas $y_{dc}$ represents the background fluorescence observed until that point;  $H(t-\theta)=0$  for $t\leq\theta$, and  $H(t-\theta)=1$ for $t>\theta$, hence this term ensures that $y(t;\tau,D,K,\tau_i,\theta,y_{dc}) =y_{dc}$ until fluorescence levels increase past the background level.

This parametrization includes many {biological compartment models} such as~\cite{choi2011dynamic,gurfinkel2000pharmacokinetics} as special cases: They typically model $I_{p,r}$ as a sum of two or three exponentials with \textbf{real} coefficients and exponents; as shown in~\eqref{eq:3exp}, the response $y_{\exp}$ allows for complex values in coefficients and exponents, and hence can model oscillations observed in ICG time-series estimated from videos of human tissue perfusion (see also Fig.~\ref{fig:fits}).

The parameters are estimated by solving the following optimization problem:
\begin{align}\label{eq:fullcost}
\textbf{\small minimize }
	&J({ \tau,D,K,\tau_i,\theta,y_{dc}; I_{p,r}}) = \sum_t
	\Bigl( y(t;\tau,D,K,\tau_i,\theta,y_{dc}) - I_{p,r}(t) \Bigr)^2 \notag\\
\textbf{\small such that }	&D,K,\theta,y_{dc} >0 \qquad\textbf{and}\qquad  0<\tau<\tau_i
\end{align}

The objective in \eqref{eq:fullcost} is a weighted least-squares data-misfit: minimizing $J$ one finds parameters $\tau,D,K,\tau_i,\theta,y_{dc}$ such that~\eqref{eq:ICGmodel} is as close as possible to the data $I_{p,r}$. The constraints in \eqref{eq:fullcost} enforce the parameters to be strictly positive. The intuition behind the constraint on $\tau$ and $\tau_i$ is 
that $\tau_i$  captures the slow decay during the {wash-out} phase, whereas $\tau$ governs the \emph{more rapid} wash-in; a faster process has a smaller time constant.

We also include a heuristic weighting term $\slfrac{W(t)}{S_{p,r}^2(t)}$, whose two parts serve different purposes: $W(t)$ emphasizes the \emph{wash-in} phase represented by approximately the first 60 seconds of the NIR intensity time-series to avoid over-emphasizing the slower decay stage, which can last several minutes. $W(t)=W_1$ for $t\le t_0$ and $W(t) = W_1 (W_1/W_2)^{t_0/(T-t_0)} e^{-\frac{\ln(W_1/W_2)}{(T-t_0)} t}$ for $ t_0<t\leq T$, so that it equals a constant $W_1$ up until $t_0$ and then decays exponentially to a smaller constant $W_2$ at the final time $T$. The constants $W_{i}$ and $t_0$ are chosen experimentally (see Section~\ref{sec:exper-valid}). $S_{p,r}(t)$ is the (thresholded, to avoid division by zero) standard deviation of the pixel brightness computed across all the pixels within ROI $r$ of patient $p$ at time $t$, and is used as a measure of data quality.

   Problem   \eqref{eq:fullcost} can be solved by any stand-alone solver supporting box constraints. In Section~\ref{sec:exper-valid-tiss}, the trust-region reflective method as implemented in SciPy~\cite{scipy} was used. Two examples of fitted responses $y(t;\tau,D,K,\tau_i,\theta,y_{dc})$ are given in Fig.~\ref{fig:fits}.

\subsubsection{Features of fluorescence intensity dynamics}
\label{sec:feat-param-fluor}

For each patient and ROI, we obtain the corresponding time-series of NIR intensity, $I_{p,r}(t)$ for $t\in \{0,\dots,T_{p,r}\}$ and estimate six parameters $(\tau,D,K,\tau_i,\theta,y_{dc})$ as suggested in Section~\ref{sec:biophys-models-fluor}. Of those, only the $(\tau,D,K,\tau_i,\theta)$ are meaningful, as the offset $y_{dc}$ depends on the background brightness and hence on the imaging equipment and the conditions under which the data is collected, but not on the tissue itself. We further derive additional features from the five basic parameters.

It is well-known from linear systems theory~\cite{phillipsharbor2000} that $y_{\exp}$ can also be represented as follows:
\begin{gather}
  \label{eq:3exp}
    y_{\exp}(t;\tau,D,K,\tau_i)  = A_{1} e^{\lambda_{1}t} +A_{2} e^{\lambda_{2}t} +A_{3} e^{\lambda_{3}t}, \\
    A_1 = \frac{K\tau_i^2}{\tau_i^2 + \tau^2 - 2D\tau\tau_i}, \quad
    \lambda_{j} = -\frac{D-(-1)^j\sqrt{D^2-1}}{\tau}\,,\quad
    A_{j} = K \frac{  D + (-1)^j\sqrt{D^2-1}- \tau/\tau_i }{2\sqrt{D^2-1} (1 -2D\tau/\tau_i + (\tau/\tau_i)^2)}\,,
\end{gather}
where $j=2,3$ and $\lambda_1=-\frac1{\tau_i}$. Intuitively, \eqref{eq:3exp} can be split into slow and fast components: the slowest component is given by $A_{1} e^{\lambda_{1}t}$ (as per the constraint $0<\tau<\tau_i$ enforced in~\eqref{eq:fullcost}), it captures wash-out and the final decay of NIR-intensity and is effectively defined by $\tau_i$ (larger $\tau_i$ -- slower decay); in contrast, the wash-in phase is mainly governed by the fast components $A_{2} e^{\lambda_{2}t} +A_{3} e^{\lambda_{3}t}$, 
the dynamics of the second-order response.

Note that while there is a one-to-one mapping between $(A_i,\lambda_i)$ and the parameters $(\tau,D,K,\tau_i,\theta,y_{dc})$ of~\eqref{eq:ICGmodel}, fitting $(A_i,\lambda_i$) to the data $I_{p,r}$ directly may require working numerically with complex numbers, and that the $A_i$ have no clear physical meaning;
hence it is hard to give physically meaningful constraints on $\lambda_i$ and $A_i$. For $(\tau,D,K,\tau_i,\theta,y_{dc})$, this is straightforward, see also the discussion below~\eqref{eq:fullcost}.

\begin{figure}[tb]
\parbox[t][][b]{.45\columnwidth}{
	\centering
	\includegraphics[width=.3\columnwidth]{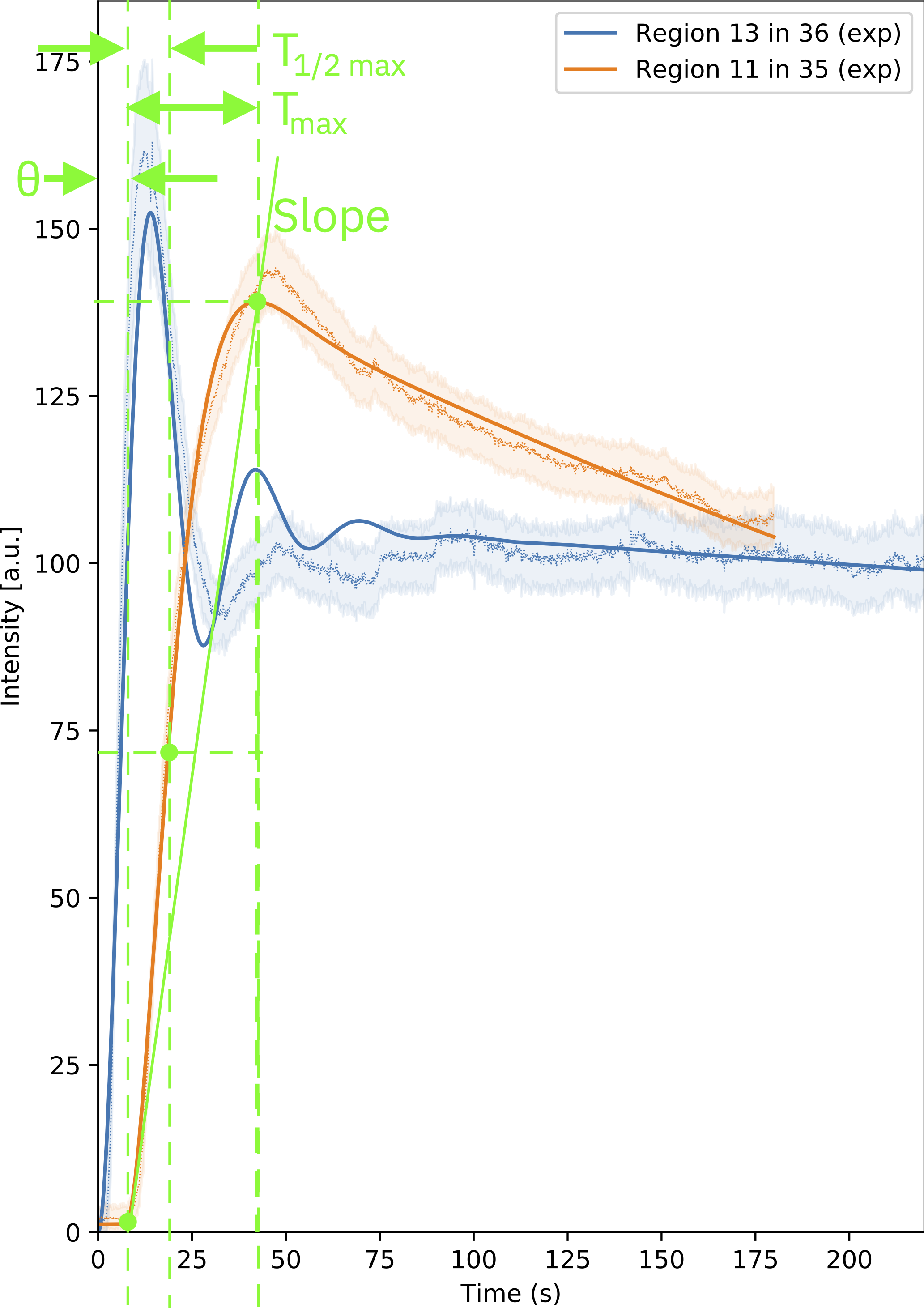}
	\caption{This figure illustrates features $T_\mathrm{max}$, $T_{1/2 \text{max}}$, \texttt{Slope}, and $T_R=T_{1/2 \text{max}}/T_\mathrm{max}$ as they are defined in~\cite{son2019quantitative}.}
	\label{fig:moson}
}\hfill
\parbox[t][][b]{.52\columnwidth}{
	\centering
	\includegraphics[width=.5\columnwidth]{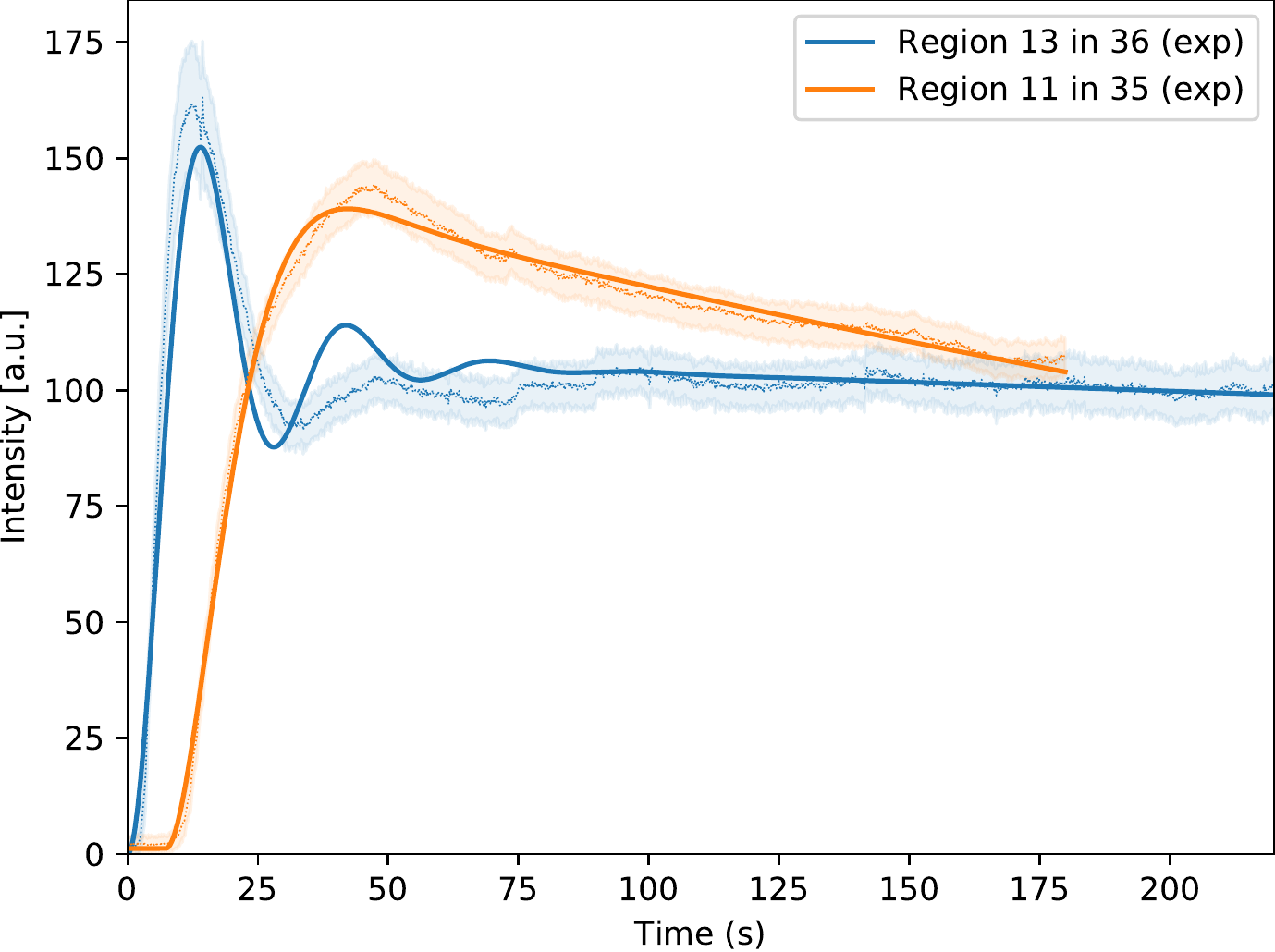}
	\caption{One profile with oscillations, corresponding to $D<1$ in~\eqref{eq:2ndorder} and complex values for $A_i$ and $\lambda_i$, $i=2,3$, in~\eqref{eq:3exp}, and one without oscillations, which corresponds to $D\geq1$ and real parameters $A_i$, $\lambda_i$.}
	\label{fig:fits}
}
\end{figure}

The real and imaginary parts of $A_i$ and $\lambda_i$, $i=1,2,3$ in~(\ref{eq:3exp}) form a set of features, which we call \emph{3EXP}. 
Another popular way of quantifying perfusion in the surgical literature is based on extracting a number of so called \emph{time-to-peak} (TTP) features $T_{\max}, T_{1/2\max},T_R$ and $\mathtt{Slope}$, which are presented in~Fig.~\ref{fig:moson}, directly from the estimated NIR-intensity time-series $I_{p,r}$. To exploit the natural synergy between surgical approaches to perfusion quantification and bio-physics we define a \emph{tumor signature} as a combination of \emph{3EXP} and \emph{TTP} features. Here we compute \emph{TTP} features directly from the \emph{fitted response} $y(t;\tau,D,K,\tau_i,\theta,y_{dc})$ after solving~(\ref{eq:2ndorder}) and getting the parameters of~(\ref{eq:ICGmodel}).
In summary, we obtain a bio-physical signature represented by the twelve features summarized in Table~\ref{tab-features}, all obtained from the \emph{fitted response}.
\begin{table}[htb]
\centering
\caption{Signature: 3EXP+TTP features}\label{tab-features}
\begin{tabular}{|c|l|}
\hline
Feature &  Bio-physical meaning \\
\hline\hline
 $  T_{\max}$	&   Time to reach first NIR intensity peak\\\hline
 $  T_{\frac12 \max}$ &   Time to reach half the above peak value\\\hline
 $  T_R$ &   Ratio of ${T_{\frac12\max}}/{T_{\max}}$ \\\hline
 $ \mathit{Slope}$&  Approximate rate of increase until NIR intensity reaches peak \\\hline\hline
 $ -\lambda_1$&  Slowest rate of decay, i.e.\ the \emph{wash-out} rate\\\hline
 $ -\Re \lambda_2,-\Re\lambda_3$& \parbox[c]{.6\columnwidth}{  Fastest and 2nd-fastest rates of decay; 
  equal if there are oscillations, see next row.} \\\hline
 $ \Im \lambda_2=-\Im\lambda_3$&   If 0, then no oscillations, else defines the frequency of initial oscillations\\\hline
 $  A_1, \Re A_2, \Re A_3,\Im A_2=-\Im A_3$ &  Coefficients of the three exponentials
 \\
\hline
\end{tabular}
\end{table}

%% file: classifier.tex

\begin{figure}[ht]
	\centering
	\includegraphics[width=.75\textwidth]{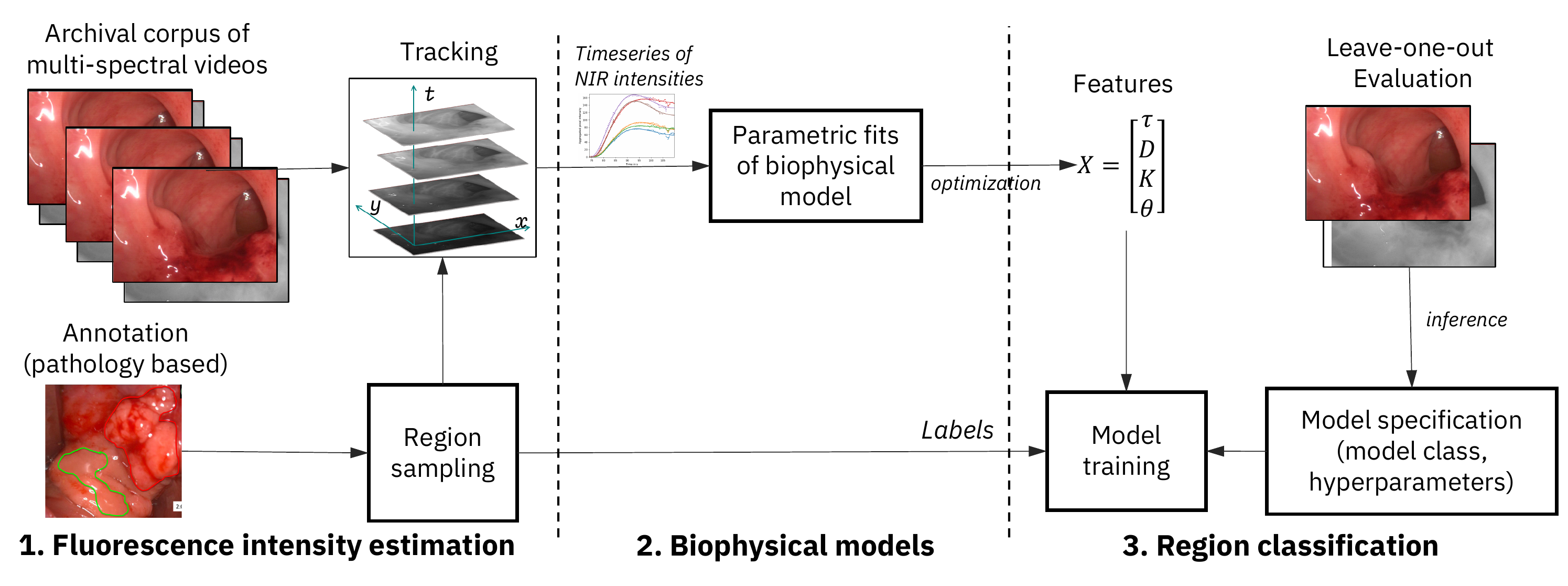}
	\caption{Process for classifier design. Each step of the pipeline was used to assemble a labelled dataset from a corpus of multi-spectral videos. A supervised classification algorithm was trained and evaluated based on a leave-one-out test framework.}
\label{fig:classifier:pipeline}
\end{figure}

The ROI classification task seeks to assign a medically relevant label to each ROI for which the signature, a vector of features described in Table~\ref{tab-features}, is available. A supervised machine learning approach is employed to perform this step. Fig.~\ref{fig:classifier:pipeline} shows an overview.
Two classification accuracy metrics are reported experimentally, one related to ROI level classification and another to patient-level correctness. The classifier is trained on a corpus of multispectral videos for which ROI annotations are provided by a surgical team and used as a groundtruth to assess \emph{ROI} accuracy. Additionally a single, per-video, case-level label is provided through post-operative pathology analysis. Case-level labels are used experimentally as a groundtruth to assess \emph{case} accuracy.

With the ROIs defined, the vector of features (see Table~\ref{tab-features}) is obtained as described in Section \ref{sec:biophys-models-fluor}. Given the number of cases available, a Leave-One-Out (LOO) evaluation scheme was employed. There are known limitations to LOO evaluation, such as higher variance for other training data, but given our cohort size we use this approach. The model is trained with ROI and case labels for $n-1$ patients and tested on the $n$-th patient by selecting random ROIs from the $n$-th video. This process is repeated $n$ times, hence each case video is used as the test video once.

The processing pipeline implements several practical quality criteria. First, quality checks on the input data extracted from videos are needed. Several filtering rules were implemented to ensure that the estimated parameters $\tau,D,K,\tau_i,\theta$ were meaningful. To ensure that the fits adequately represented the underlying data, a threshold of $10\%$ was also considered on the $L_1$ loss. ROIs that failed these quality checks were returned with a `No Prediction' label.

We consider three class labels: `normal', `benign', and `cancer'. For ROIs meeting the quality criteria above, the classifier returns class probabilities for each label. We experimented with two variants of the classifier. A three-class model, where each ROI was labelled as one of the three classes, and a two-class model where only suspicious ROIs were classified as either cancer or benign. The two-class model performed significantly better and results are presented for this case in Section~\ref{sec:datas-eval-results}.

A noisy-OR aggregation from the set of predicted ROI-level classifications is used to derive a patient-level label. Specifically, for a case with $n$ ROIs, $c$ of which are predicted to be cancerous, the case level class probability is derived as $P(\text{'cancer'}) = \frac{c}{n}  (1-p_\text{fp}) + \frac{n - c}{n}  p_\text{fn}$ where $p_\text{fp}$ and $p_\text{fn}$ are the false positive and false negative rate  estimated from a validation sample.

Significant patient-level variations were observed in the data.
To mitigate this in the two-class model, a set of healthy ROIs are used as a reference for feature values. The classifier is trained with features regularized using the healthy reference. This normalizes the features across patients. All features were split into two classes: `rate features' (e.g. $\lambda_1, \Re\lambda_2, \Re\lambda_3, \mathit{Slope}$), and `absolute features' (e.g. $T_{\max}, T_{12 \max}, T_R$). `Rate' features were normalized as a ratio, e.g. $\lambda_1^{cancer}\mapsto \slfrac{\lambda_1^{cancer}}{\lambda_1^{healthy}}$, and `absolutes' were subtracted from the reference value.

%% file: experiment.tex
\paragraph{\bf The data.}
The proposed signature was evaluated by performing ROI classification on a dataset of 24 patients (11 with cancer) comprised of 24 multispectral endoscopic videos with annotations of suspicious and healthy ROIs (Figure \ref{fig:classifier:pipeline}). ROI annotations were provided by a surgeon, and pathology findings per patient (normal, cancer, benign) were given from post-operative pathology analysis.

To evaluate classifier performance, several ROIs were selected at random for each patient(ranging between 11 and 40 ROIs per case) defining a total of 526 samples with labels. For each of these ROIs, NIR-intensity time-series were estimated as described in Section~\ref{sec:endosc-moti-comp}. The length of the time-series focused on the initial \emph{wash-in} period: time-series lasted between 100 and 300 seconds for each ROI. Each time-series was next assigned a signature of features described in Table~\ref{tab-features} by solving ~\eqref{eq:fullcost} using the \texttt{curve\_fit} functionality from the \texttt{optimize} package of SciPy~\cite{scipy} by means of the \emph{trust-region} reflective method (\texttt{'trf'}) and taking $\tau<100$, $\tau_i>150$, $W_{1}=10$, $W_2=1$ and $t_0=100$. 

Several ROIs from the resulting dataset were discarded as quality thresholds were not met. Specifically, seven ROIs were discarded as the time series was too short, several others if damping coefficient $D$ was unrealistic. The time constant $\tau$ on 32 occasions was too large and the fit L1 error exceeded 10\% on nine occasions. After discarding all these cases, the resulting dataset had 435 samples from 20 patients (8 with cancer) with 12 features per sample as shown in Table~\ref{tab-features}. The resulting dataset of ROIs was relatively balanced in terms of outcomes: ``benign'' (n=198), ``cancer'' (n=124), and ``normal'' (n=181).
\paragraph{\bf Classifier.}
We experimented with several combinations of feature sets, filtering rules, and machine learning algorithms. The best performing feature set was based on \emph{3EXP}-features combined with \emph{TTP}-features (Table~\ref{tab-features}). Several standard classifiers from the Scikit-Learn package \cite{scikit-learn} were tested. The \emph{ensemble gradient boosted tree method} was found to perform the best. This model was further refined using a grid search to fine tune the hyper-parameters. An example of a classification result is given in~Fig.~\ref{fig:p18}, panel C.
%
\begin{table}[!htb]
	\centering
	\caption{Evaluation results comparing several classifiers for the same features. The last two rows show the best performing pipeline on different feature sets.}
	\label{tb:classifier:results}
	\input{result.tex}
      \end{table}

\paragraph{\bf Results.}
Table \ref{tb:classifier:results} shows results for a set of tested classifiers. For the best performing pipeline, using the gradient boosted classifier, mean LOO accuracy score for ROI-based correctness was 86.38\%, i.e. the fraction of ROIs correctly predicted in unseen patients. Case accuracy to be 95\% (19 out of 20 cases), i.e. the number of unseen patients correctly classified. The best performing pipeline has a 100\% cancer sensitivity and 91.67\% specificity for patient-level classification. The results strongly suggests that the signature, defined by \emph{3EXP- and TTP-features} is discriminant.

We also show a comparison of feature sets. \emph{TTP}-features offer a slightly higher performance on ROI classification accuracy compared to the exponential features (3EXP). However, 3EXP has a 15.7\% point improvement in case accuracy, 27.8\% point improvement in sensitivity, and a 8.34\% point improvement in specificity (a difference of two patients).

%% file: result.tex
\begin{tabular}{llrrrr}
\toprule
Features & Model & Mean ROI accuracy & Case Accuracy & Sensitivity & Specificity \\ \midrule
3EXP + TTP & Nearest Neighbors &              71.49\% &        65.00\% &      50.00\% &      75.00\% \\
& SVM (RBF)         &          76.60\% &        65.00\% &      25.00\% &      91.67\% \\
& Gaussian Process  &            68.51\% &        60.00\% &       0.00\% &     \textbf{100.00\%} \\
& Decision Tree     &         85.53\% &        90.00\% &     100.00\% &      83.33\% \\
& Random Forest     &           78.30\% &        65.00\% &      50.00\% &      75.00\% \\
& Naive Bayes       &           48.09\% &        50.00\% &      37.50\% &      58.33\% \\
& QDA               &          54.04\% &        55.00\% &      62.50\% &      50.00\% \\
& \textbf{XGBoost } &               \textbf{86.38\%} &        \textbf{95.00\%} &     \textbf{100.00\%} &      91.67\% \\ \bottomrule
3EXP & XGBoost  &           76.95\% &        85.71\% &      77.78\% &      91.67\% \\
TTP &  XGBoost    &             82.55\% &        70.00\% &      50.00\% &      83.33\% \\

\bottomrule
\end{tabular}